\documentstyle[epsfig,times,12pt]{article}
\def\be{\begin{equation}}
\def\ee{\end{equation}}
\def\br{\begin{eqnarray}}
\def\er{\end{eqnarray}}

\begin{document}
\title{Science In Trauma}
\author{Bhag C. Chauhan{\footnote{chauhan@cftp.ist.utl.pt}}\\
\it Centro de F\'{i}sica Te\'{o}rica das Part\'{i}culas (CFTP)\\
Departmento de Fisica, Instituto Superior T\'{e}cnico\\
Av. Rovisco Pais, 1049-001, Lisboa-PORTUGAL}
\date{\today}
\maketitle
\begin{abstract}
Quantum theory has been proved as an outstanding mystery in modern science. The predictions of science have turned out to be probabilistic. The principle of determinism has failed. For systems like weather, earthquakes, rolling dices etc... and of course human behavior it has proved impossible, for science, to describe a state of the system accurately for a long time into the future. Moreover, modern cosmology has to rely on philosophical assumptions. It is argued -- by taking into account of the views of learned scientists and philosophers -- that modern science can never explain everything and it is a hard fact to discover the ``Theory Of Everything''. On the other hand, modern scientific technology have damaged our environment and posed serious threats for the future development of science. All these facts and results put a big question-mark (?) on the grass-root level working of science. In fact, all scientific researches are based upon ordinary sense perception, which keeps the outer physical universe as a separate entity, that is something quite independent of the observer. Basically, it is the observer -- the knower -- which makes perception possible. Astonishingly, the knower -- human mind -- is not included in our scientific theories. One should not forget human-being (human mind) is a part of nature and an essential component of our observations, so it's inclusion in scientific theories is must in order to recover modern science from the drastic state of trauma. 
\end{abstract}
%%%%%%%%%%%%%%%%%%%%%%%%%%%%%%%%%%%%%%%%%%%%%%%%%%%%%%%%%%%%%%%%%%%%
\newpage
%\baselineskip=2\baselineskip
\section*{Motivation and Objectives}
The spectacular success of science in the last century has assented people to consider it to be a self-contained worldview independent or inclusive of its entire philosophical foundation.
Advancement of science has definitely implicated the enormous physical comforts but at the same proportion a boom of miseries. A soaring level of human happiness and the resulting endless lusty desires have given birth to several problems. And in order to fight those problems man has developed nuclear weapons, polluted the green-earth-environment and nourished a criminal world. 

On the other hand, there are various difficulties in our theories which prevent us to understand the true picture of nature and reality. 
Along with the scientific knowledge, we have also gathered speculations, debates, and confusions. 
As we are extending the radius of our scientific knowledge, so have been increasing the circumference of our ignorance and the truth is becoming more and more dispelled. 
After struggling for several years a desponded scientist, Albert Einstein, who have a number of pioneering contributions\footnote{He gave a new vision to the dynamics of moving bodies by postulating special theory of relativity and the understanding of gravitational force as space-time curvature; discovered mass-energy relation, Brownian motion, photo-electric effect etc...} to the development of modern science, uttered: ``I used to think when I was young that sooner, or later all the mysteries of existence would be solved and I worked hard. But now I can say that the more we know, the more our existence turns out to be mysterious. The more we know, the less we know and the more we become aware of the vastness... Science has failed in de-mystifying existence, on the contrary it has mystified things even more.''
John Bahcall -- a leading solar-neutrino physicist and astrophysicist -- once proclaimed in a commencement address to the Physics and Astronomy graduates of the University of California, Berkeley \cite{r3}: ``Science is unpredictable and fun''.

All the crisis that humanity and modern science are facing today, clearly show that there is something wrong in our conventional way of thinking, the way we understand nature and our relation with it. Certainly, we need to understand our minds and learn to eliminate the causes which promote human toward destruction, so that a flower of peace and compassion could blossom on the green globe. That's why a scientific contemplation of human mind is essential. 
By taking into account of the views of learned scientists and philosophers it has been argued that there is a serious need to include the properties of human mind -- the real observer -- in our scientific theories in order to understand the truth sitting at the heart of nature. In fact, in our scientific methods we exclude the real observer from the observation. This is ridiculous: A big mistake that we are doing and must be corrected!  

In this article, I discussed obvious consequences (Sections 1-5) of the flawed approach of our thinking and a natural limitation of perception (Section 6), which have made science miserable and tarnished its bright future.
In fact, I have explored a fundamental problem in the scientific foundation of perception of the objective reality and nature, which have bogged down modern science into a drastic state of trauma. A diagnosis has been proposed with some thought experiments to recover science from its traumatic state in Section 7. Evolution of life into a ``New Paradigm'' -- consciousness revolution -- is shown in the end: Section 8.

The fundamental theory of matter -- ``Quantum Theory'' -- is plagued by several conceptual difficulties is shown in Sec. 1. In Section 2 it is presented that ``Determinism'' -- the soul of science -- is a hard fact to establish in complex classical systems, and it is replaced by postulating another principle called ``Uncertainty Principle'', for quantum systems. Section 3 shows that modern cosmology -- a theory which describe the origin and evolution of the universe -- does not have any scientific basis for the questions about its origin. A secret dream of scientists for centuries -- ``Theory of Everything'' -- seems not to be fulfilled is conjectured in section 4. There persists always a limitation in perception of the outside world, which obstructs the observer and thereby science from the true picture of reality is argued in Section 5. However, Section 6 introduced another obstacle ``Repercussion of Nature'', which is putting a gigantic threat to the human survival and thereby to the further development of science.
   
\section {Quantum Theory: A Mystery}
Quantum theory grew out of a series of anomalies in the picture of matter and light offered by classical physics -- in particular associated with black-body radiation, the photo-electric effect, and the need to devise a model of the atom consistent with the newly discovered sub-atomic particles. Without quantum physics, we are unable to explain the behavior of solids, the structure and function of DNA, super-conductivity, properties of super-fluids, and burning of stars etc...
There is no doubt that quantum theory has been one of the most profound discoveries of the $20^{th}$ century development of science. In fact, this theory has become dramatically successful in order to explain the experimental results, which were, otherwise, impossible to understand in the classical formalism. 

Despite the extraordinary successes, quantum theory has been plagued by conceptual difficulties. The debate about the relation of quantum mechanics to the familiar physical world continues. It is not at all clear, what this theory is about and what does it, in fact, describe? \cite{sgold}.   
From its inception the theory had has a ``measurement problem'' with the troubling intrusion of the observer \cite{brfk}\footnote{The linearity property of Schroedinger equation advocates the multiple outcomes and forbids to pick up the unique result in an experiment. There is no mechanism in the theory by which the predicted multiple possibilities become the single actuality after the involvement of observer at the act of measurement.} in experiments.  
An Irish physicist J.S. Bell\footnote{Who became well known as the originator of Bell's Theorem, regarded by some in the quantum physics community as one of the most important theorems of the 20th century.} has quoted in his book, "Speakable and Unspeakable in Quantum Mechanics" \cite{bell}: ``... conventional formulations of quantum theory, and of quantum field theory in particular, are unprofessionally vague and ambiguous. Professional theoretical physicists ought to be able to do better.''

Albert Einstein was not at all comfortable with the foundation and working of quantum theory, despite the important role he had played in the development of this theory (he was awarded the Nobel Prize for discovering the photo-electric effect). 
It is a general conviction among the scientists that Niels Bohr\footnote{A Danish physicist who received Nobel Prize in 1922 for his services in the investigation of the structure of atoms and of the radiation emanating from them.} (founder of Copenhagen interpretation of quantum mechanics) vanquished Einstein in their famous, decades-long, debate \cite{debate}. Nevertheless, till the end of his life, Einstein continued to pretend that perhaps the quantum mechanical description is not the whole story.
Erwin Schrodinger, one of the founders of the quantum theory and who is also known as the father of wave function\footnote{A continuous function that contains all the measurable informations about the quantum particle.}, was one of the most acerbic critics of the theory. He ultimately found this theory as impossible to believe.

\vspace{0.4truecm}
{\it Mysteries, puzzles and paradoxes} in Quantum Mechanics:

\vspace{0.3truecm}
\begin{small}
\noindent
However, the Schrodinger equation is perfectly linear, propagates continuously in time, but collapses discontinuously when a particle interacts with a classical system at the event of measurement. In fact, there is no dynamical description for the {\it collapse} of the wave function.

\vspace{0.4truecm}

\noindent
A quantum system is described with a complex wave function ($\psi$) which is an abstract entity, but whose squared value ($\psi^2$) represents its physical properties. This gives a probability distribution for where discrete particles may be found once the wave function is collapsed by an act of observation.

\vspace{0.4truecm}

\noindent
Quantum particles can have spooky connections: According to theory, they can communicate over vast distances in an instant, which gave rise to the famous EPR paradox \cite{EPR} and Bell's theorem \cite{bell1}. This ghost action violates the principle of the limitation of the velocity of light in relativity theory and the principle of causality. 

\vspace{0.4truecm}

\noindent
There is a profound relationship between measurement and reality, where reality depends heavily on the measurement techniques. Observation would create a different kind of reality than what existed independently. In other words, reality existed in a different way while under observation than it did in itself.

\vspace{0.4truecm}

\noindent
According to the ``Principle of Superposition'' the Schroedinger's Cat inside a box \cite{scat} is neither {\it dead} nor {\it alive}, but a superposition of these two states. The wave function thus contains the superposition of all possible states of a system until it is observed. 

\vspace{0.4truecm}

\noindent
A quantum particle can behave as a {\it wave} as well as a {\it particle}; e.g., in photo-electric effect it shows its particle nature whereas in a double slit experiment it behaves like a wave.

\vspace{0.4truecm}

\noindent
It is fundamentally impossible to measure the key physical quantities, in certain pairs, e.g., position and momentum, simultaneously to any desired degree of accuracy. Attempts to increase the precision of one measurement, result in less precise measures of the other member of the pair: ``Principle of Uncertainty.''
\end{small}

\vspace{0.6truecm}

There were several attempts to falsify this theory on conceptual and experimental grounds, e.g., Albert Einstein with the collaboration of Boris Podolsky and Nathan Rosen proposed a gedanken experiment [EPR Experiment]\cite{EPR} as an attempt to show that quantum mechanics was somehow not complete and that the wave function does not provide a complete description of physical reality. However, they left open the question of whether or not such a description exists.
J.S. Bell proved mathematically through an inequality, famously known as ``Bell's Inequality''\cite{bell}, that quantum mechanics does violate special relativity by allowing instantaneous interactions across even the cosmological distances.  In another attempt, Erwin Schroedinger fabricated a thought experiments \cite{scat}: {\it Cat-in-a Box}, where the future of a cat paradoxically depends on the random decay of a radioactive atom. Astonishingly, according to quantum theory, the hapless cat is neither dead nor alive but in a state of superposition of the two possibilities, before to be seen actually -- which is ridiculous and hard to swallow.

\section {Determinism is Killed}
``Science is deterministic; it is so {\it a priori}; it postulates determinism, because without this postulate science could not exist'', stated by Henri Poincar\'{e} (1854-1912) -- a French mathematician and physicist. So, if there is a scientific law working anywhere in the universe, determinism must be there. According to another French mathematician (and astronomer) -- Pierre Simon Laplace (1749-1827) -- the entire future course of the universe is laid out as a consequence of two deterministic factors: 1) The laws of nature. 2) The state of universe at any moment of time.
Universe is nothing but a collection of numerous phenomena. Anything happening around us, in the universe, is controlled by a certain set of physical laws. These laws determine the birth, life, and death of all the events of universe. So, with the total understanding of these physical laws, one can predict the future of an event very well in advance: This is the beauty of science, as Henri Poincare quoted. 

In contrary, the facts suggests that the claim of determinism has turned out to be a failure in science. For classical systems like weather, earthquakes, rolling dices, turbulence of fluids, oscillations in the populations of organisms, fluctuations of populations in complex ecosystems, etc... and, of course, human behavior, it has proved impossible scientifically, to describe a state of the system accurately for a long time into the future. Instead, probabilities can be derived to describe a state of the system which might happen in future. 

Besides suffering from several difficulties, quantum theory has postulated the {\it ``Principle of Uncertainty''}\footnote{According to this principle, if we take an electron, for example, and try to determine its position by using electromagnetic radiation, since electrons are so small, we need radiation of very short wavelength to locate it accurately. However, shorter wavelengths correspond to higher energies; so as we use the radiations of higher energy, the momentum of the electron is altered. In this way, the attempt to determine the location accurately, changes the velocity of the electron. Conversely, techniques for accurately measuring the velocity of the electron will leave us in ignorance about its precise location.} which has, of course, killed determinism -- the soul of science. Albert Einstein bothered much with this principle. He believed in an universe of deterministic principles and once said: ``God does not play dice''.
This principle was highly controversial in the beginning but later the experimental results have supported it.
Werner K. Heisenberg (a German Physicist who discovered the principle) himself, took a more radical view on claiming that this limitation is a property of nature rather than an artifact of experimentation \cite{heisnbrg}. Also, Niels Bohr pointed out that the indefiniteness of the quantum theory is due to the inherit nature of atoms at the atomic level.     
 
\section{Modern Cosmology on a Philosophical Base} 
Sir Arthur S. Eddington (1882-1944) -- a British astrophysicist (the most distinguished astrophysicist of his time) -- who was one of the first to appreciate the importance of Einstein's theories of special and general relativity, and published a treatise on the subject, once said: ``The universe no longer looks like a thing but like a thought''. 

As far as the origin and evolution of our universe -- cosmology -- is concerned, the most popular theory in modern science is the {\it ``big-bang theory''} \cite{bigb}. According to this theory there was nothing before the big-bang and all the space-time must have originated there and then (``t=0''). No matter/ energy could exist before this bang, as there was no space and time for it to be in. The theory further describes that this universe evolved from a dense, nearly featureless hot gas and that is expanding and cooling continuously. 

Scientific evidences strongly support that the universe had a definite beginning a finite amount of time ago and also prove that the early universe was very hot and that as it expands, the gas within it cools. There are three important observations strongly supporting  the big-bang model: 1) The expansion of the universe observed in 1929 by Edwin Hubble. 2) The abundance of the light elements H, He, Li (according to the theory these light elements should have been fused from protons and neutrons in the first few minutes after the big-bang). 3) The discovery of the Cosmic Microwave Background (CMB) radiation. The theory claims that the CMB radiation is the remnant heat leftover from the big-bang and the frequency spectrum of the CMB should have a blackbody radiation form. This was indeed measured with tremendous accuracy by an experiment on NASA's COBE satellite. The recent Wilkinson Microwave Anisotropy Probe (WMAP) mission reveals conditions as they existed in the early universe by measuring the properties of the CMB radiation over the full sky \cite{WMAP}.

Although, this theory has passed some scientific tests, there are still many more trials, which it must undergo successfully.
In the context of a recent test of this theory, John Bahcall writes \cite{r6}: ``I am happy that the big-bang theory passed this test, but it would have been more exciting if the theory had failed and we had to start looking for a new model of the evolution of universe''. 
In fact, there are many domains of modern cosmology which are far from being settled. The theory is silent about what banged, why it banged, or what happened before it banged. Despite its name, the big-bang theory does not describe the {\it bang} at all. The biggest problem of the big-bang theory of {\it the origin of the universe is philosophical} -- perhaps even theological -- what banged and why it banged!  

The philosophical base of the theory stands as embarrassing situation for the scientists. Robert Jastrow -- the first chairman of NASA's Lunar Exploration Committee -- himself admitted \cite{r5a}: ``Astronomers try not to be influenced by philosophical considerations. However, the idea of a universe that has both a beginning and an end is distasteful to the scientific mind''. 
To avoid this initial difficulty the idea of {\it singularity} was introduced in which the universe expands from a singular point and collapses back to the singular point and repeats the cycle indefinitely \cite{r5}. The idea was appreciated to avoid the philosophical, rather theological, base of the theory, but the available experimental evidences indicate that this type of oscillating universe is a physical impossibility. The facts and recent results suggest the geometry of the universe is flat and will expand forever \cite{r8,WMAP}. So, the attempts behind this idea to avoid philosophical or theistic beginning of the universe all fail \cite{r9}.

The philosophical origin of the big-bang is difficult to rule out even in the current attempts that are being made through a highly speculative theory of unification of quantum mechanics with gravity: ``Quantum Cosmology''. It must be noted that the meaning of ``t=0'' is highly contextualized by the assumptions and limitations of big-bang theory. In the alternative theories like quantum cosmology, they may well address the problems like ``t=0'' but the underlying philosophical ideas about space, time, matter and causality, far from being eradicated, will re-emerge in new and distinctive patterns and which will lead to further questions.

\section{TOE Project at Halt}
According to Occam's Razor, {\it when multiple explanations are available for a phenomenon, the simplest version must be preferred} \cite{occamr}. This principle argues for the economy of understanding nature: The logical description of a vast range of physical phenomena from a few basic principles, rather than the memorization of a large number of isolated facts or formulae. Such economy is the strength of modern analytical science.

Scientists have a secret dream to expound nature in the {\it simplest version}. They want to explain all phenomena in the universe with the minimum number of particles interacting with a single interaction. Search for such a Theory Of Everything (TOE) is like the quest for the Holy Grail in the Middles Ages.

TOE is a beautiful contemplation of theoretical physics and mathematics that fully explains all the known and unknown -- everything in entire universe including life -- with a single unified equation.
Search for such a theory has started from the idea proposed by Isaac Newton. According to him, one great theory might exist that would link all the other known theories and this grand unified theory (GUT) would be able to describe everything including life in the entire universe.

Science has traversed a long way since the time of Newton, and other physicists, including Albert Einstein, began to realize this beautiful idea of unification. This idea became more popular after the revolutionary work of James Clerk Maxwell (1831-1879): The first theoretical unification of the two physical phenomena -- electricity and magnetism -- into one all-encompassing framework. 
The next great step was the success of Quantum Electrodynamics (QED) theory (the integration of electromagnetism and quantum mechanics)\footnote{This landmark work in the direction of GUT earned Richard Feynman, Julian Schwinger, and Sin-itiro Tomonaga the Nobel Prize for physics in 1965.}.
On the same lines, the unification of electromagnetic and weak nuclear forces known as Electro-Weak theory (EW)\footnote{In 1979, Sheldon Glashow, Abdus Salam, and Stephen Weinberg were given the Nobel Prize for this work.} took place. 

There are continuous endeavors to find the most promising road to a GUT. Scientists have created some theories that can by themselves form a GUT, namely {\it string theory, supergravity, and loop quantum gravity}.
They want to see this theory (GUT) as an unification of the two great bastions of twentieth century physics: 1) Theory of the very big ({\it general relativity theory}) that describes the large scale structure in the universe. 2) Theory of the very small ({\it quantum theory}) that studies the microscopic structures.

Although, the idea of unification seemed quite rewarding, yet the several difficulties at theoretical, experimental and phenomenological level have faded away the hope of realization of this elegant dream: What once seemed very near on the horizon may be further off than imagined.
Much of the difficulty in merging these theories comes from the radically different assumptions that these theories make on how the universe works. On the one hand, in conventional GUTs like SU(5) physical particles exist in the flat space-time of special relativity, whereas on the other hand in general relativity space-time is curved and that changes by the motion of mass. 

Noticing that a class of GUT quantum theories proposed in 1980's and later \cite{gut} couldn't pass even the first test in the laboratory: In 1999, Superkamiokande experiments reported that they had not detected proton decay as predicted by the GUTs \cite{sk99}. Also none of the generic predictions of these theories -- the existence of topological defects such as monopoles, cosmic strings, domain walls etc... -- has been observed yet. As a result, not a single such quantum theory is currently universally accepted. Obviously, at this stage of debates and confusions, unification of all the four interactions is extremely difficult.

On the other hand, the very complexity of Einstein's general relativity was first noted by himself as leading to a very serious impediment on its further development. In fact, after publishing his famous paper in 1916, he conceded that this arose from the mathematical difficulties involved in the complexity of its nonlinear coupled equations and their huge number of terms.                          In 1952 he expounded it as an acute frustration: ``The generalization of the theory of gravitation has occupied me unceasingly since 1916.''
   
On the lines of principle of economy -- Occam's Razor -- I believe TOE must be simple, elegant, pure, and perfect theory, which is clean and spotless, and {\it devoid of uncertainties in measurements and predictions}. Or to put it differently, TOE is a complete deterministic theory. In this way, quantum theory fails even the prime eligibility condition for the holistic project of grand unification. Noticing that this essential quality of determinism is, unfortunately, entombed in the foundation of the theory itself as ``Principle of Uncertainty''\cite{copenhgn}. 

In a response of a question: ``Can science explain everything?'' the Nobel Prize-winning theoretical physicist, Steven Weinberg, replied \cite{r1}: ``Clearly not. There certainly always will be accidents that no one will explain, not because they could not be explained if we knew all the precise conditions that led up to them, but because we {\it never} will know all these conditions''. 

\section{Hammer of Nature}
There is no doubt about the fact that the progress of science has been much better in the $20^{th}$ century and very likely this progress will be intensified in future. Nonetheless, there is a risk never reaching a climax of science, where our understanding of the whole universe including life would be perfect. If we just overlook the above discussed difficulties that modern science is waging with and be highly optimistic (assuming that the climax of science is not far off), even then there is another problem ready to interrupt the further progress of science and from which, it seems, the science can never escape: the {\it repercussion of nature}. 

Advancement of science has changed all the spheres of life: social, economical, political etc... Our thoughts, plans, and policies have become subjective, selfish, and short-visioned. The development of physical comfort has raised the level of human happiness. To be happy one needs a lot of luxury. One's desires have been increased excessively. As a result, there has been an incredible growth in various vices; like anger, lust, ego, greed, and selfishness etc... In order to enjoy these immoral habits, one negatively benefited the advancement of scientific technology -- developed nuclear weapons, polluted the green earth environment and nourished a criminal world. 
In the words of an American Civil-Rights Leader -- Martin Luther King, Jr.: ``Our scientific power has outrun our spiritual power. We have guided missiles and misguided men''. 
According to Einstein: ``Concern for man himself must always constitute the chief objective of all technological effort -- concern for the big, unsolved problems of how to organize human work and the distribution of commodities in such a manner as to assure that the results of our scientific thinking may be a blessing to mankind, and not a curse''.

Modern science and technology have damaged our environment and posed serious threats for our future. Rather than harnessing, nature has been over-tortured by the development of industries on an extensive scale and other technological implications.  
Remember that for every action, there is an equal and opposite reaction. We should not forget that mother nature is vindicating to each of our vicious acts on it. One can see this as
an insidious impact that global warming is putting on the climate change and thereby on the individual species as well as the entire food chain \cite{dte}. Also as the deteriorating quality of food, fruits, and vegetables with the obsession to increase farm productivity.
  
There is something which is bothersome at this stage of high alert: The way the nuclear weapons are being developed, the way the environment is being polluted and the resources of earth are being exploited, the way the restlessness of the world is growing are, of course, not going to wait for the science to flourish up to its climax. Eventually, if all these things keep on growing on such an exponential rate, there is no certainty whether that culmination would ever be reached, because before that perfectness is attained, the modern civilization might be destroyed by natural calamities, dangerous weapons and radiation hazards, and several other insidious impulses caused by ecological imbalances. 

So, one can see, there is a clear paradox: Progress of science and technology at a hallucinating pace in the one hand may offer a beautiful world with longevity and leisure etc... and on the other their byproducts that are generated in the equal proportion may destroy the whole social fabric so that the beautiful world may be out of reach.

In summary, we must be awake and vigilant of the admonishing of mother nature; and remind ourselves, that nature may fight back sometime in an unpredictable and terrifying way -- A colossal {\it hammer of nature} is prompt to dismantle our modern civilization and, of course, science. 
And there will be nobody to ask the question: {\it ``Is the climax of science, nearby?''} The man of genius, Albert Einstein, once warned about such a probable danger as a byproduct of modern scientific technology in future, with a quote: ``The fourth world war will be fought with  bows and arrows.'' 

\section{Fallacy in Perception} 
``No phenomenon is a phenomenon until it is an observed phenomenon'' -- Niels Bohr once promulgated in the context of quantum mechanics. It is a known fact that to make a perception possible, three basic elements must be present: first, the {\it observer}; second, the {\it observable}; and third, the {\it connection} between the two. In order to understand nature, we consider the human body -- a frame with the five sense organs and a brain -- as the observer, an event happening in nature as the observable, and the electromagnetic wave spectrum (EMS) as the connection. 

It is well acknowledged that the human body is having a limited ability of perception of the outside material world. Because of its limitations, this body is also called by some philosophers as: ``A Limited Instrument''. It has five senses of perception: vision, hear, taste, smell, and touch. Nature has granted a narrow band of the electromagnetic spectrum, for connection with the observables, to this body: a visible band of light and a part of the infra-red region\footnote{Although, ``sound'' and ``odor'' first propagate through the air pressure, yet the message is finally conveyed to the brain via an electromagnetic signal.}. It is true that with the development of science, we could expand our vision -- communicate better with the observables -- and thereby understand nature better. The EMS window of perception is being broaden by our scientific knowledge -- the scientifically detected (known) part of the spectrum extends from a short wavelength of cosmic rays to the long wavelength of power induction ($\lambda~\approx~10^{-13}~c m~-~10^{5}~cm$). 
 
It is reasonable to speculate that the EMS is unbounded; what lies beyond, on either side of the known region, we are not aware of. At this stage the communication between observer and object/reality is weak; we can have a stronger and stronger communication as we explore the more and more parts of the EMS. We can only have the full information of the object/universe and thereby the knowledge of ultimate reality of nature, only if we could explore the whole unbounded EMS. This means that the picture of our universe seen through the infinite unbounded EMS must be the ultimate one and at this state the communication is said to be "total". So, the "total" communication can be achieved by using the knowledge of the unbounded parts of spectrum that lie beyond either side of the currently known range.
 
We have seen that the human body has limitations and is not perfect. The understanding beyond either side of the unbounded EMS, even with the prospective enhanced scientific vision, up to any finite extension will always be incomplete. In other words, augmenting the knowledge of this spectrum up to any finite extension will always be infinitely small as compared to the infinite whole of it. Meaning that the communication is again within a limited band of frequencies -- through a window of the EMS. This shows that there persists always an error in perception.
In this way, it seems impossible to explore the knowledge of this vast unbounded EMS and thereby establish a "total" communication.  
Clearly, the present efforts to attain the complete knowledge of our universe, life, and entire existence seems not going to be fulfilled, in any way, with such a fallacy in perception. In this way, the limit of perception would certainly truncate the human rational approach and hence set a concrete limit to the further growth of science.

\section{\it A Diagnosis} 
Modern scientific advancement has influenced all the sectors of our day-to-day life including our thoughts and culture. It is evident that along with the enormous physical comforts, mental restless and all the problems at personal, social, and global levels that we face today are also related to our scientific understanding of nature: The way we understand the working of mother nature and our relations to it.  
So, there exists essentially a crisis of understanding of our own minds and nature -- a crisis of perception and cognition. 

Although, we know that our thoughts and emotions do influence our brain chemistry and other biological activities, yet for no significant reasons we don't treat them in the definition of reality.
In fact, all scientific researches are relying upon ordinary sense perception, a prejudice posed by classical physics, which keeps the outer physical universe as a separate entity, to be an independent existence, that is something quite independent of the observer. So, basically, we are separating the real observer from the observation; this approach is famously known as ``Cartesian Partition''. 
However, it is well acknowledged that the classical physics is an incomplete understanding of nature. So, our scientific researches based upon Cartesian partition approach must be incomplete or erroneous.  

In principle, a serious flaw is resting at the grass-root level working of science. There is a need to understand fully the working of this objective method of scientific studies, which is only relying upon the sense perception of human body. To make a perception possible there must be a subject --the knower-- who can observe a phenomenon or an event with the help of a connecting principle. 
{\it In fact, it is not the physical part of human brain which acts as the observer (the knower) and makes the perception possible, but there exists a subtle playback entity; a consciousness being -- ``Mind''. The human mind is the doer, the observer which interprets the messages collected from outside by the brain with the help of sense organs and instruments.} It makes a person or scientist to recognize or refute the existence of an object or a phenomenon. 

Notice that the most creative physicists have always emphasized that human consciousness (mind) is at the foundation of the scientific method behind physics.
According to Eugene Wigner -- American Physicist, Nobel Laureate: ``The next revolution in physics will occur when the properties of mind will be included in the equations of quantum theory''. Luis De Broglie -- who proposed the idea of the wave-nature of particle -- said: ``The structure of the material universe has something in common with the laws that govern the working of the human mind''. Erwin Schroedinger felt deeply that human mind is a sole constructor of all the observations and quoted as: ``Our picture of the world is, and always will be, a construct of the mind''. 

To a first approximation, Cartesian approach is fine and workable, as evidenced by the success science since its birth, but if the target is to find the ultimate answer to all questions which a conscious being can think of, then one has to eliminate such separation of observer and object from the scientific study of nature\footnote{The relationship of mind and matter, which eliminate Cartesian partition, has already been explored in literature e.g. see David Bohm \cite{bohm2}.}. 
Astonishingly, the science working in the heart of nature -- quantum theory -- couldn't avoid the influence of the observer at the act of measurement -- famously known as the ``measurement problem''.
This acute ascendancy of observer in the measurement process experimentally confirms the prevailing importance of the functioning of ``human mind'' -- the real observer -- in the definition of reality\footnote{However, one may argue that these two things have nothing to do with each other, but I do believe that they are somehow correlated and kins. Certainly, this argument deserves a more attention and investigation.}.

Now, the biggest {\it challenge} for the scientists is, namely, to include the real knower (the doer) -- human mind -- in their theories. However, this ``call'' is not restricted only to the scientific community, but for all of the people who are wise and of good will, because (as we have seen) all the problems in the world, which have made it miserable, are also rooted in the functioning of our own minds\footnote{If we, all humans on the green globe, become apprised of the functioning of our minds, we would be able to really understand the root causes of our problems and difficulties, and feel the pains and miseries of all animates around. Such that, we would become simple enough to accommodate the life, love, and freedom of each one living being. In this way, an ecological revolution -- a true religiosity -- would emerge.}. Such that, there are strong possibilities of interdisciplinary co-operation; the experts from all other disciplines of life: human psychology, health, ecology, economics, education, administration, and politics etc... may join forces (in any means) and contribute substantially for the holistic cause.

In the following subsections, I have examined some concrete and tentative properties of human mind and nature, which could help and serve as a tool in order to incorporate human mind, the observer, into the scientific theories:

\subsection{\it Objectivity of Human Mind}
Science is working to fulfill the ``Laplacean dreams'' of a mechanical universe. It is an objective study of matter and its interactions; based on a tacit assumption that information about a physical system can be acquired without influencing the the system's state. However, the "information" is regarded as unphysical, a mere record of the tangible, material universe essentially decoupled from the domain governed by the physical laws.  

The Cartesian approach of science finds it extremely difficult to include an entity, in the theory, which is just a subjective experience. If human mind is purely subjective in nature, as per general prejudice, then it would certainly be impossible to contemplate it scientifically. On the other hand, the objectivity property of mind can dramatically simplify the understanding of it and thereby its inclusion in scientific theories. 

Metaphorically, human mind can be understood as a lake of water. The surface of lake is the dividing line between sleeping and waking state of consciousness. Below the surface is the subconscious and unconscious mind, and on the surface is the conscious state of mind.
I believe that human mind is subjective only in its superficial layers which retain and defend the individuality of a person. However, the objectivity is in the whole remaining part of it, as because all humans are alike in the deep down, having the similar qualities: quest for truth and happiness, an urge to be alive, and seek justice, sense of guilt, love and compassion, and kindness etc...). For example, if you ask a terrorist; why he was shedding blood of innocent lives. His answer is simple and straightforward: ``I don't like to do but am forced by the circumstances.'' This clearly shows the objectivity of mind below the superficial ``terrorist-mask''. (For more elaboration see the gedanken experiment 1.)

Traditionally, using statistical analysis, the three factors genetic, environmental, and the product of both components have been recognized as responsible for all the behavioral variations.
It has been accepted since a long time by the biologists that the genes, the environment, and the interaction between them orchestrate the human behavior \cite{gene}. 
Which shows that only these three factors contribute to shape the personality of a child and offer a shallow individuality to him. So, in this way, the subjectivity of human mind at the superficial levels (although not well-understood yet) can be scientifically contemplated through neuro-studies, genetics and in behavioral sciences. 

\subsection*{Gedanken Experiment: 1}
 The mind of every new-born child is like a {\it formatted and blank} compact disc (Let's name it metaphorically as: ``Mind-Dics''.) produced by the same company (nature) through a certain franchised firm (parents). To call it {\it formatted}, I mean that there already exist some genetic and biological instructions; like, a certain mental level and some specific traits, e.g., cry for food, sucking of nipple, excretion, breathing, and blood circulation etc... all of those necessary to carry life. The ``Mind-Disc'' is called {\it blank} means that it is like a clean slate, what-so-ever you want to print on it, you can. So, let's take the mind of a new-born child as a fresh sample to test. As child grew up social  factors start shaping his life. It is a noted fact that the parents, teachers, and society are the dominant factors responsible for an innocent child to become a criminal or a gentleman in the later life. Evidently, this example, to a far extent, supports the objectivity of human mind and thereby establishes that determinism is working on the most part of it. The other factors influencing the human behavior are genetic and biological (as a format of the mind-disc set off by nature). If we could contemplate them scientifically, a total objectivity of human mind can be established.

\subsection{\it Probability to Determinism}
It is quite hard to believe that any random process, we see around, which shows up the probabilistic feature {\it in practice}, is deterministic {\it in principle}. So, to understand that a random process is a deterministic in principle, it is desirable to test an obvious and prime example of probability theory, first: ``Tossing of a Coin''(See Gedanken Experiment: 2).  

There are several other random phenomena, which are more complex and non-trivial to prove as deterministic. For example, the principle of determinism seems not working in chaotic systems; like, turbulence of fluids, oscillations in the populations of organisms, fluctuations of populations in complex ecosystems, and weather patterns etc...but this indeterminacy could be because of our incomplete understanding of the scientific laws, which are governing their complex dynamics. Interestingly enough, these systems have been found extremely susceptible to changes in the initial conditions. Apparently, there is no reason to assume that only accidents are prevailing there in the evolution of such nonlinear systems.
Chaos theory supports a strictly deterministic philosophy of nature, although in epistemic limit.

Extending this argument further in the same lines, there should not be any trouble in beholding the evident notion that the whole microscopic particle world is deterministic at the heart. Nevertheless, the indeterminacy in the measurement of observables at the microscopic level has been discovered\footnote{Quantum Mechanics tells us that microscopic reality is irretrievably fuzzy.}. Now, let's try to understand the meaning of the phrase ``indeterminacy in the measurement'' clearly. Notice that the uncertainty principle of quantum mechanics simply tells that it is hard to know e.g., the definite position and velocity of a electron at one instant of time. {\it It doesn't tell us that the electron, at any instant of time, does not possess a definite position and velocity}. ``[Physicists] ... convert what is not measurable by them into unreal and the nonexistent''\cite{Adler}.

\subsection*{Gedanken Experiment: 2}
Tossing of a coin is fun! In our day-to-day life we toss a coin (as a symbol of randomness) just to qualify the chance in a game. This is really funny, because here we exploit the limitations of our perception in order to enter into an amusement world. The limitations of our sense perception make it impossible to predict the real outcome for this process: either "Head", or "Tail". In fact, the tossing of a coin is a deterministic, and not a random process. Only because of some unknown variables; like movement of air and the way it is flipped, we loss determinism and fail to predict the result. But, if we perform this experiment in a place where there is no air and flip the coin with a machine so that we would know exactly how it was flipped, then we could predict precisely the outcome. Which clearly establish determinism in a probabilistic process. 

\subsection{\it Power of Human Mind}
If we recourse through the history of evolution of life on earth, we will know that amongst all the other creatures it is only human mind that has gone through a prodigious mutation. Human mind has been continuously earning power through the billion years process of evolution. 
It has been asseverated that human mind has power to achieve anything; radiant health, a loving relationship, become wealthy and successful, accomplish the true happiness and fulfillment. 
It is also a known fact that the secret of all the achievements in one's life is in the tuning of his mind to; which can be achieved through dedication, devotion, and hard work. In fact, all these factors increase the subtlety and the power of mind. It is just like what happens in a convex lens: The concentration of all the rays from sun at the focal point increases the intensity of light at that point and burn the sheet of paper. 

Amazingly enough, dreams are somewhat bizarre tools justifying the enormous power of mind. 
For most of us, dreaming is something quite separate from normal life; a dream world is totally separated from our normal world of reality. Nevertheless, there are noted facts about dreams which show a cogent relationship of the two worlds of reality that can not be denied.
There are observed facts that human mind can vividly perceive things in dreams that we can't observe otherwise in the state of awake. 
The dreams in which the dreamer knows at the time that he is dreaming, are called "Lucid Dreams" \cite{lucid}. Lucid dreaming has been noted an incredibly powerful information tool, since the dreamer can then take control of their dream and work through, or confront the issues they feel they need to deal with. 

Today there are various fascinating studies which ratify this fact as a grand source of creativity\footnote{Creativity in physiological terms, means physical healing and regeneration; in emotional terms, it means creating attitude changes that favor the establishment of inner harmony; in the mental sphere, it involves the synthesis of new ideas and scientific discoveries.}:

Famously, Friedrich Von Kekule revolutionized organic chemistry with his findings about the structure of the benzene molecule. Surprisingly enough, all details came to him in a dream. One night he dreamed of a serpent eating its tail and when he woke up, he got a flashing insight into the circular rather than linear nature of benzene molecule. The circular structures of certain processes of growth are fundamental in molecular biology. 

The Russian chemist Dmitri Mendeleev -- known as father of periodic table -- saw in a dream a table where all the elements fell into place as required. After working for years to discover a way of classifying the elements according to their atomic weights, he got the formula in a weird style. One night in 1869, he fell into bed exhausted after devoting many long hours to the problem. Later that night, he saw the periodic table in his dream and upon awakening, he immediately wrote down the table just as he remembered it. Amazingly, the periodic table of the elements first brought forth, after a mere readjusting the position of a single element, was a fundamental discovery of modern science \cite{Mendlf}.  

In an another example of creative dreaming, Elias Howe, the inventor of the sewing machine, had worked on his idea for years before attaining success. He had solved the secret in a formidable dream -- cold sweat poured down his brow, his hands shook with fear, his knees quaked and he got the answer. All this was so real to him that he cried aloud. What he needed was a needle with an eye near the point! \cite{Howe}.

Golf player, "Jack Nicklaus" saw the solution to a problematic golf swing in a dream. After winning a number of championships, he had found himself in an embarrassing slump, but dramatically he regained his championship form seemingly overnight \cite{Nick}. Nevertheless, Nicklaus confessed that he has felt foolish admitting that really happened in a dream.

Ann Faraday in ``Dream Power'' reported a similar case in which a gynecologist discovered how to tie a surgical knot deep in the pelvis with his left hand while dreaming \cite{Farad}.

Another whimsical tool that exhibit the power of human mind in a weird style is the ``intuition''. The intuitive experiences (where mind illuminates just as a ``click'' of clock) are not just restricted to the domain of mysticism; they work beautifully for science as well. 
In general, it has been noticed that while doing science, as one worked very hard for days or weeks to understand the concept following along a logical path to explain an {\it anomaly}, and it did not work. Finally, he got frustration and wanted to give up; got relaxed or spontaneously involving himself with something else. Then eventually all of a sudden, the right idea popped up and got the answer: The solution was just right there. So, what that worked finally was the power of mind!   

The confession of Nicklaus suggests that there may be others with similar experiences who have never mentioned the source of their inspiration. Most of the scientists are skeptic to admit the very fact that the {\it intuition} worked remarkably in their inventions of great ideas. Only the uncommon minds, like Einstein, can dare to: ``I maintain that the cosmic religious feeling is the strongest and noblest motive for scientific research''. 
Certainly, any idea, whatever big it may be, comes up traversing through the vast realm of human consciousness with monstrous churning and agitating, and there is no way to separate it from the influence of mind itself\footnote{So, at some point a scientist must agree upon the deep ascendancy of his mind on the understanding of nature and its principles.}.

We have seen in these examples, the dream and intuitions play a key role in the illumination phase of the creative process. 
There are several undocumented examples showing the enormous power of perception of human mind through dreams and intuitions, and some other similar practices \cite{undoc} through which all the relevant informations about the object are collected. 
I believe, in such experiences and practices the subtlety and fine tuning of mind is somehow achieved and limitations of the observer are eliminated\footnote{This apparently shows that human mind has ``power of elimination'' of the limitations of the observer.}, and a "total" communication with the target/objective reality is established. 

\subsection{\it Total Communication}
It is a well known fact that out of five senses of perception as one sense is dropped, the knowledge about the reality of nature gets truncated and life becomes rather difficult to carry. For example, the life of a blind man is a bit miserable because his knowledge of the outside world is restricted. He can not observe visible light but only a part of infra-red region. As a result, he is unable to understand the beauty of colors, landscapes, sunset, moon, and stars etc... and above all, he faces lots of difficulties to accomplish the tasks of daily life.  

In fact, it is our poor communication with the objective world that makes us unable to understand and enjoy the grace of the objective reality.  
As an experiment if we take an infrared photograph of our study room, we will not be able to make out the picture: ``A complete mess!'' We can not identify the scattered objects (chair, table, books and other things), in the room, in the way we see them normally with the visible light. This would be something similar like an attempt locating things in a dark room. 

On the other hand, with the help of the scientific instruments, we see better and know more about nature by improving the communication; because a wider spread of the EMS is being utilized. With the scientific vision, we can now see the region of the universe which was practically inaccessible to us before. For example, images of our universe taken in gamma rays have yielded important information on the life and death of stars, and other violent processes in the universe. Moreover, x-ray images of our sun can yield important clues to solar flares and other changes on our sun that can affect space weather. The technological implications of the known part of the EMS have made us able to look deep inside the objects and the associated phenomena. In other words, we have got a much better picture of our universe as a whole. 
So, more we explore the EMS, or the more we improve the communication, the better we know about the intimate secrets of universe and the reality of nature. 

\begin{figure}
\centerline{\includegraphics[width=17pc]{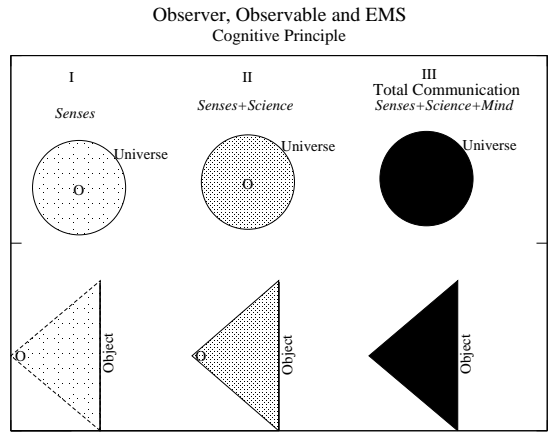}}
\caption{A schematic depiction of the phenomenon of "total" communication (state III). State I represents a weak communication (only through the senses) while state II  with a bit more shaded region shows a better communication achieved with the help of scientific methods.}
\end{figure}

The "total" communication requires a vision through the infinite extent of the EMS, which is an ideal situation and seems unattainable, because an infinite entity is incomprehensible. 
Nevertheless, I believe that it is possible to establish the "total" communication with the objective reality, once the observer is completely merged with the object/universe (the complete dark shaded state (III) shown in the figure 1, where the observer "O" is totally absorbed with the object and information). In other words, one has to go beyond the duality and just unite with the observable completely through the glue of informations, and become the only ``One''. This is, in principle, the highest state of awareness. In this state of unification, mind can explore all the informations from the object or universe. There have been undocumented incidents and histories \cite{undoc} which suggest that such type of perceptions and cognitions are, in practice, possible. 

\section{\it Evolution into A New Paradigm}
History of evolution taught us that there have always been occurring major leaps in the process of evolution at several points starting from big-bang, formation of galaxies and stars, solar system and earth, emergence of life and consciousness to the delirious development of science in the last century. 

Many scientists estimate that life on earth originated between 3.5 and 4.0 billion years ago. Once life arose on earth, a number of major events occurred during biological evolution \cite{MSS}. The geological record indicates that the oldest micro-fossil of a prokaryote (Cyanobacteria) is roughly 3.5 billion years old. The first eukaryotes are thought to have appeared around 2 billion years after the first prokaryotes. 
In general, eukaryotes include animals, plants, fungi and protists. The land plants and land fungi are thought to be existed about 480 to 460 million years ago, and there is some evidence that plants colonized the land around 600 to 700 million years ago. 
Fossil records show that new species evolved from the multicellular organisms within relatively short time periods. The rapid increases in species diversity, for example, {\it the Cambrian explosion}, show that there has been a number of evolutionary explosions through the history of life. 

The archaeological survey shows that the beginnings of a reflective consciousness emerged decisively at the time roughly 35,000 years ago. During this period the numerous developments, as found in the excavations and investigations, were occurring in stone tools, burial sites, cave art, and migration patterns; a first awakened human culture was born in these glimmerings of personal and shared awareness. 
There happened a dramatic change in our view of reality and human identity at about roughly 10,000 years ago when our ancestors shifted from a nomadic life to a more settled livings in villages and farms; and then followed by a rise of city-states and the beginnings of civilization at roughly about 5,000 years ago.
 
A more recent revolution in the human awakening, which is clearly visible through the existing vast literature, museums, and all the pathways of our life happened roughly 400 years ago with a radical dynamism and materialism of the industrial era. The scientific revolution gave a totally different awareness to the human understanding of reality -- all aspects of life have vividly changed with it, including the work that people do, the ways they live together, how they relate to one another, and how they see their role in society and place in the universe.

The humanity's prevailing paradigm is changed again by another radical worldview, which was kicked off in the beginning of $20^{th}$ century, with the emergence of a new vision of matter and universe. The modern concept of matter in subatomic physics from {\it quantum theory} and the new concept of space-time from the {\it theory of relativity} are totally different from the one, of which we were traditionally used to. These new explorations have changed our conception of the universe as a whole including life in it.

Today, our perceptual paradigm may be again undergoing one of such rare shifts.
In the light of difficulties plagued to our scientific theories, crisis at the global, social, and personal levels and the emergence of theories of self-organization in living organism and ecological theory, we have before us the possibility of a very different understanding of humanity's relation to nature \cite{capra}. In the view of above, there are strong chances that we may witness an emergence of a ``New Paradigm'' in the near future.
So, according to the history of evolution, a transformation of our conventional world-view, our thinking; the way we understand nature and humanity as a whole is quite possible.  

\section*{Prospectus}
Quantum theory has been proved as an outstanding mystery in modern science. The predictions of science have turned out to be probabilistic. The principle of determinism has been entombed in the foundation of theory. For the classical systems like weather, earthquakes, rolling dices etc... it has proved impossible, for science, to describe a state of the system accurately for a long time into the future. Moreover, modern cosmology has to rely on philosophical assumptions. 

On the other hand, the advancement of scientific technology has rendered enough physical comfort, which has raised the level of human happiness. The endless human desires have given birth to various problems. In order to fight those problems man has developed nuclear weapons, polluted the green-earth-environment and nourished a criminal world. As a result modern civilization is full of terror and hatred, and human existence on the green globe is in peril.

At this moment of high alert, there is a serious need for introspection -- to realize: what we have done, what we are doing, and what is going to happen. All the problems we face today at the global, social, and personal level and the difficulties that modern science is confronted with are rooted directly or indirectly to our thoughts. There is something wrong with the way we understand nature and its relation to mankind.  
It has been argued that there is persisting a serious flaw in our basic understanding of nature. {\it All the scientific researches keep the outer physical universe as a separate entity, that is something quite independent of the observer. The observer -- human mind -- is not included in our scientific theories, despite the fact that human mind is a part of nature and an essential component of our observations.} 

The difficulties faced by science and the entire humanity, have forced us to expand the worldview in order to include human mind in the definition of reality. 
Indeed, the quantum physics experiments have knocked the door of a new paradigm through the troublesome intrusion of observer -- human mind -- in the act of observation. It is also argued that the prevailing role of observer, as suggested by the quantum physics experiments, is a physical proof of the idea floated in Section 7 as a diagnosis.        
Several measures have been suggested and elaborated with thought experiments, as an attempt to resolve this problem. The objectivity property of human mind has been brought into light. As a result of which, it would be dramatically simpler to comprehend it in the scientific theories. 

Nevertheless, the inclusion of human mind is a great challenge for the scientists. It is asserted that there are strong possibilities that a fruitful collaboration of the experts from all disciplines of life could spark and may facilitate the accomplishment of the holistic cause and recovering modern science from the drastic state of trauma. Notice that this type of radical change in the human understanding of nature and objective reality is also supported by the several leaps happened in the history of evolution.  
The paradigm-shift has the potential to dramatically transform our view of reality, identity, social relationships, and human purpose. 

\subsection*{Upshot} 
The diagnosis proposed as an inclusion of human mind in the scientific theories is {\it new}\footnote{In the sense that it points out a serious flaw in the foundation of science related to the conventional methods of perception and cognition.} and the realm of human mind is {\it not-yet-understood} clearly. It is nevertheless of scientific importance because of the tangible evidences shown in Section 7 and 8. Once some concrete steps are made in this direction, the solutions to the various problems related to the difficulties and growth of modern science, and peace and prosperity of humanity would start showing up.

\subsection*{Acknowledgments:}
I am grateful to Jo\~{a}o Pulido and Gregory Moreau for the careful reading of the manuscript and useful suggestions.


\begin{thebibliography}{99}
\bibitem{occamr} Occam's Razor: A principle attributed to the 14th century English logician and Franciscan friar, William of Occam that forms the basis of methodological reductionism, also called the principle of parsimony. This has become a basic perspective for those who follow the scientific method.
\bibitem{undoc} Undocumented: There have been several undocumented practices, more common in the yore, in which a specialized person collects desired informations about the past and future life of the subject just by fixing his mind to. These techniques were mainly used for the purpose of healing from ailments. Such methods are still applied mostly in the far-flung regions of globe where the influence of scientific advancement is weak.
\end{thebibliography}
\end{document}